# *In-situ* TEM observation of preferential amorphization in single crystal Si nanowire

Jiangbin Su[1,2], Xianfang Zhu[1*]

1. *China-Australia Joint Laboratory for Functional Nanomaterials & Physics Department, Xiamen University, Xiamen 361005, PR China*

2. *Experiment Center of Electronic Science and Technology, School of Mathematics and Physics, Changzhou University, Changzhou 213164, PR China*

\* Corresponding author. E-mail: zhux@xmu.edu.cn

**Abstract:** The nanoinstability of single crystal Si nanowire under focused electron beam irradiation was *in-situ* investigated at room temperature by transmission electron microscopy technique. It was observed that the Si nanowire amorphized preferentially from the surface towards the center with the increasing of electron dose. In contrast, in the center of the Si nanowire the amorphization seemed non-uniform and much more difficult accompanying with rotation of crystal grains and compression of d-spacing. Such a selectively preferential amorphization as athermally induced by the electron beam irradiation can be well accounted for by our proposed concepts of nanocurvature effect and energetic beam-induced athermal activation effect, while the classical knock-on mechanism and the electron beam heating effect seem inadequate to explain these processes. Furthermore, the findings revealed the difference of amorphization between Si nanowire and Si film under focused electron beam irradiation. Also, the findings have important implications for nanostability and nanoprocessing of future Si nanowire-based devices.

# 1. Introduction

It has been reported that semiconductor Si nanowires have potential applications in nanodevices owing to their special electronic and optical properties [1,2]. However, the intrinsic structural instability of Si nanowires due to their nanosize effect, especially due to their nanocurvature effect, may have negative effects on their performance and lifetime in service. Thus, it becomes imperative and crucial to study the structural instability (called as nanoinstability or nanoprocessing in general) of Si nanowires especially under external excitation such as energetic beam irradiation. In this sense, the delicate, energetic field emission focused electron beam in transmission electron microscope (TEM) has been proven to be uniquely versatile and powerful [3-5]. Up until now, the focused electron beam in TEM has been mainly applied to create holes, gaps and other patterns on an individual Si nanowire, or to weld an individual Si nanowire with metal nanowires to form semiconductor-metal junctions [6]. In addition, Dai *et al* [7,8] have reported the crystalline-amorphous transition of Si nanowire and the subsequent plastic bending as induced by electron beam irradiation; Han *et al* [9,10] have studied the axially-extending and bending of Si nanowire and related amorphization process as-induced by external tension force along with electron beam irradiation; He *et al* [11] have realized the shear-driven dynamic amorphization process in silicon crystals using *in situ* high-resolution TEM. In spite of these, details and mechanism of the amorphization especially the preferential amorphization in Si nanowires as purely induced by electron beam irradiation have not been revealed. In addition, the amorphization of Si nanowire is expected to be greatly different from that of Si film [12-15] due to the unique nanocurved wire shape. Thus, the fundamental nanoscience underlying the above irradiation-induced nanoprocessing phenomena in nanowires such as the nanocurvature effect [16,17] and the energetic beam-induced athermal activation effect [16,18] have not been explored yet.

With the above considerations, in this paper we particularly study the amorphization of single crystal Si nanowire under focused electron beam irradiation. We report an intriguing preferential amorphization from the nanowire surface as athermally induced by the focused electron beam irradiation. In contrast, in the center of the Si nanowire the amorphization seemed non-uniform and much more difficult accompanying with rotation of crystal grains and compression of d-spacing. These findings have important implications for structural stability and engineering of the future Si nanowire-based device. More importantly, the findings demonstrate that the current knock-on mechanism [19] and the electron beam heating effect [6,20,21] are inadequate to explain these processes but our proposed nanocurvature effect and energetic beam-induced athermal activation effect obviously dominate the processes.

## 2. Experimental

Single crystal Si nanowires synthesized by chemical vapor deposition method were firstly treated in 5% hydrofluoric acid for 10 minutes, then dispersed and diluted in 99.7% ethanol, and finally collected on holey carbon grids (with large holes in the electron-transparent film) to prepare TEM specimens. With such a treatment, it is expected that the oxide layer on the surface of the wires was fully removed. An *in-situ* irradiation on the treated wire was subsequently carried out at room temperature in a field emission TEM (FEI Tecnai F30) operateing at 300 kV with a focused beam of about 90 nm in diameter and beam current density of about $1.5 \times 10^2$ A/cm$^2$. The beam was perpendicular to the wire axis and focused on the center of the wire. Also, the irradiation was performed on a settled segment of an individual nanowire protruding into the open space of the holes in carbon film so that local structure transformation or deformation of the wire is not hindered by any support within the holes. It is estimated that the penetration depth of such electrons exceeded the TEM sample thickness or nanowire diameter and the energy deposition rate would be uniform over all the materials in the sample [22]. During the observation or taking a picture, the beam was spread to an around 100 times weaker intensity so that the corresponding irradiation effect can be minimized to a negligible degree and at the same time the image contrast can also be improved. In the experiment, several characterization techniques such as TEM, high resolution TEM (HRTEM), fast Fourier transform (FFT) and energy-dispersed X-ray spectroscopy (EDX) were applied for a detailed analysis of the shape, structure and composition of the wires.

## 3. Results and discussion

As shown in Fig. 1(a), both HRTEM and FFT images illustrated that the Si nanowire we used here was almost of a well-defined (111) single crystal structure (as indicated by Area A) before irradiation except a very thin amorphous layer of 3-4 nm in thickness enveloped on the surface of the wire (as indicated by Area B). Since the surface of Si nanowire specimens has been fully treated *via* hydrofluoric acid, any possible original oxide layers on the wire surface formed during its high temperature growth should be completely removed. Thus, the observed amorphous layer here on the surface could only be a newly and preferentially formed amorphous Si layer after the chemical etching. Indeed, our further quantitative EDX analysis indicated this tendency. As further addressed in the following discussion part, with a positive activation volume [16], the amorphization transformation can preferentially occur from the surface of a solid where the surrounding open space can be provided for the volume expansion during the amorphization. Especially for the nanowire here, its high surface energy associated with its high surface nanocurvature [16] can provide an additional driving force for such a preferential amorphization.

After irradiation for 308 s, as shown in Fig. 1(b), both HRTEM and FFT images demonstrated that the central part of the Si nanowire amorphized non-uniformly, for example, with an increase of amorphization degree from Area C to E. If we regard the less-amorphized areas (such as Area C and D) as crystal grains, the sizes of them are different, from several nanometers to ten nanometers or more. More intriguingly, such grain fragments are still (111)-orientated but their directions are different, Areas C and D for example. We further measured the angle between the directions of Areas C and D, and got the value of 48.7 degree. This value is greatly inconsistent with that determined by crystallographic symmetry, in which the angles between any two of (111), (-1-1 1) and (1-1 1) orientations should be 109.5 or 70.5 degree. So it indicates that there was an adjustment or rotation of the (111) grains under electron beam irradiation. This is probably due to the surrounding space change or stress unbalance caused during the above non-uniform amorphization. Furthermore, the d-spacing values of the (111) grains in the central part of the Si nanowire were also different, all of which were slightly smaller than that before irradiation as shown in Fig. 1(a). It is expected that the amorphization in the center of the Si nanowire accompanying with an at least 1.8% [23] volume expansion would slightly contract or compress the d-spacing of the grains. Therefore, the amorphization in the center of the Si nanowire is incomplete, non-uniform and seems to be difficult. Note that the volume expansion of 1.8% is an ideal status that there is not any vacancy in the amorphized material with a continuous network structure. However, the real volume expansion should be larger or even much larger than 1.8% due to the existence of vacancies in amorphous material. In this work, the volume expansion is estimated to be about 6%, which is much larger than the ideal value probably due to a further production of vacancies in the amorphous Si as caused by electron beam irradiation.

While on the surface of the wire, as shown in Fig. 1(c), there were three layers could be differentiated clearly (i.e., Layer F, G and H) according to their crystal structures or image contrast. In details, Layer F was a layer nearly non-amorphized and still of crystalline structure (see the HRTEM and FFT images). Layer G and H, however, were almost amorphous with obviously different image contrast (see the corresponding HRTEM and FFT images). Further EDX analysis shown in Fig. 2 illustrated that the main components of Layer G (also covered by Layer H right above) and H were silicon and carbon respectively. The amorphous carbon layer (Layer H) probably came from the residual organic gas in the TEM chamber or the supporting organic films which decompounded under focused electron beam irradiation and further deposited onto the surface of the Si nanowire [24]. The Layer G of amorphous Si was about 8 nm, which was much thicker than the initial amorphous Si layer in Fig. 1(a). It was resulted from a further and complete amorphization from the wire surface during the irradiation. Relative to the incomplete and non-uniform amorphization in the center of the Si nanowire, the amorphization on or near the wire surface seems to be much easier, preferential and uniform. Note that the near-surface Layer F of crystalline structure seems larger than the crystal grains in the wire center. This can be attributed to the different amorphization behaviors in the wire center and on the wire surface. For the specified Layer F, the side close to Layer G is suffering from a uniform amorphization from the surface toward the center and thus presents a continuous interface. Although Layer F is getting thinner and thinner, it still needs more time to get amorphized fully. Hence, we can see a long strip-like or thin layer of crystalline Si close to the amorphous Si shell at an intermediate time such as the 308 s in Fig. 1(c). In contrast, for the central part of wire, it is suffering from a non-uniform, random amorphization and thus the as-formed non-amorphized areas or crystal grains exhibit fragments-like shape. In this way, the long strip-like Layer F near the wire surface seems larger than the fragment-like grains in the wire center.

According to the conventional views, the main mechanisms of interaction between energetic electron beam and solid materials are knock-on mechanism [19] and electron beam heating effect [6,20,21]. However, the knock-on mechanism and related simulations neither are fully consistent with, nor can offer a full explanation for, the experimentally observed wire structure changes, especially the above preferential amorphization on the surface and formation of a local coaxial structure during the irradiation. This is because the electron irradiation energy for knock-on displacement of Si atoms is about 440 keV [10], which much exceeds the value of 300 keV applied in this work. More importantly, the existing theories such as the knock-on mechanism and related simulations were at the first place built on consideration of the nature of equilibrium, symmetry, periodicity and linearity of bulk crystalline structure or its approximation, whereas the energetic beam-induced nanophenomena are intrinsically of non-equilibrium, amorphous and non-linear nature. Moreover, the simulations are of roughness and the atomic movements on the scale of atomic bond lengths are difficult to be observed with current TEM techniques, and thus a direct comparison of the simulated atomic structure changes of the wires with the experimental results is impossible. On the other hand, due to the extremely high ratio of surface to volume of a nanowire at the nanoscale, the electron beam irradiation is expected to heat the specimen by no more than a few degrees [7,19,22,25-27], and the dominant irradiation effect should be athermal although a direct and precise measurement of temperature of a nanoscaled TEM specimen has been rather difficult so far.

In fact, our previous work on the energetic beam irradiation-induced structural instabilities and processing of low dimensional nanostructures (LDNs), such as nanocavities in Si [16,28-31], carbon nanotubes [32,33] and amorphous $SiO_x$ nanowires [22,34], has proven that our proposed nanocurvature effect [16,17] and energetic beam-induced athermal activation effect [16,18] are universal concepts and applicable in the prediction or explanation of energetic beam irradiation-induced nanophenomena including the amorphization of LDNs. For the amorphization of a solid, the open space which can act as the positive activation volume [16] is firstly required for the volume expansion during the amorphization. Especially for a nanowire here, there is sufficient open space surrounding the wire surface but little open space in the center of the nanowire. It is thus expected that the amorphization on the nanowire surface should be much easier and preferential. Besides, as demonstrated in the following, an additional driving force on the nanowire surface which is resulted from the nanocurvature effect would further dominantly contribute to the preferential amorphization.

For the nanocurvature effect on a nanowire, we can suppose that, similar to the particle case [16,17], when the diameter of a nanowire reduces down to nanoscale and can be comparable to its atomic bond length, a positive nanocurvature on the highly curved wire surface will become appreciable, as shown in Fig. 3(a). Such a positive nanocurvature would cause an additional tensile stress on the electron cloud structure of surface atoms which would lead to a dramatic increase in surface energy. This dramatically increased surface energy could provide a driving force for the atoms to migrate or escape thermodynamically and thus cause a strong tendency of amorphization transformation for a crystalline nanowire especially on the wire surface. In this sense, the intrinsic nanocurvature effect of Si nanowire is thus regarded as one of the key factors contributed to the preferential amorphization on the wire surface.

Although the positive nanocurvature can provide the driving force for the atoms to migrate or escape thermodynamically, a further assistance from external excitation such as energetic beam irradiation is still needed to lower the energy barrier and realize kinetically the amorphization. In the case of energetic electron beam irradiation in TEM, we can assume that, when the beam energy deposition rate of the incident energetic beam becomes very fast, there is no enough time for the deposited energy to transfer to thermal vibration energy of atoms within a single period of the vibration, and thus the vibration of atoms would lose stability or the mode of atomic vibration would be softened [16,18]. The as-induced atom instability or soft mode can thus athermally activate the structure transformations such as the amorphization of Si nanowire not only on its surface but also in its center, as illustrated in Fig. 3(a). In this way, as driven by the effects of nanocurvature and beam-induced athermal activation, the Si nanowire amorphized preferentially on the wire surface (see Figs. 1 and 3).

Note that the electrons of 300 keV can penetrate the TEM sample thickness or the Si nanowire diameter easily with a typical energy loss of only several keV. The influence of sample thickness on the uniformity of beam energy deposition rate should thus be negligible. Furthermore, as schematically illustrated in Fig. 3, the diameter of focused electron beam is much larger than that of the Si nanowire (90 nm *vs*. 67 nm). Although a focused electron beam often exhibits a Gaussian-like intensity profile, the intensity is regarded to be uniform within the central zone of the beam which can fully cover the wire diameter (see Fig. 3). Therefore, the beam energy deposition rate would be uniform over all the materials in the nanowire sample. In this sense, the preferential amorphization on the surface of Si nanowire is only attributed to the high nanocurvature on the wire surface along with a large positive activation volume. In contrast, as reported in Refs. [12-15], it was found that Si film amorphized preferentially in the central zone of the focused beam. As illustrated in Fig. 3(b), the Si film is flat without detectable nanocurvature on the film surface. Accordingly, there is no nanocurvature effect working on the preferential amorphization. Instead, the focused electron beam has a higher flux or beam energy deposition rate in the central zone of the beam relative to that at the edges of the beam (see Fig. 3). As a result, the effect of such Gaussian-like intensity profile of incident energetic beam determine the amorphization process of the Si film. That is, the Si film would amorphize preferentially and quickly in the central zone of the beam while difficultly and slowly at the edges of the beam. In this sense, the nanocurvature- or intensity-dependent preferential amorphization behavior is probably the greatest difference of amorphization between Si nanowire and Si film under focused electron beam irradiation.

## 4. Conclusions

In this work, the focused electron beam irradiation-induced athermal amorphization of single crystal Si nanowires was investigated at room temperature in a field emission TEM. With the beam spot diameter larger than that of the Si nanowire, the local wire segment within the irradiated area demonstrates a selectively preferential amorphization on the surface during the irradiation. Such an amorphization can be well accounted for by our proposed concepts of nanocurvature effect and energetic beam-induced athermal activation effect whereas the classical knock-on mechanism and the electron beam heating effect seem inadequate to explain these processes. Furthermore, the findings revealed the differences between amorphization of Si nanowire and that of Si film under focused electron beam irradiation. Also, the findings have important implications for nanostability and nanoprocessing of the future Si nanowire-based devices.

## Acknowledgements

We thank Prof. Deren Yang at State Key Laboratory of Silicon Materials, Zhejiang University for supplying the Si nanowire samples. This work was supported by the NSFC project under grant no. 11574255, the Science and Technology Plan (Cooperation) Key Project from Fujian Province Science and Technology Department under grant no. 2014I0016, and the National Key Basic Science Research

Program (973 Project) under grant no. 2007CB936603.

**Figures**

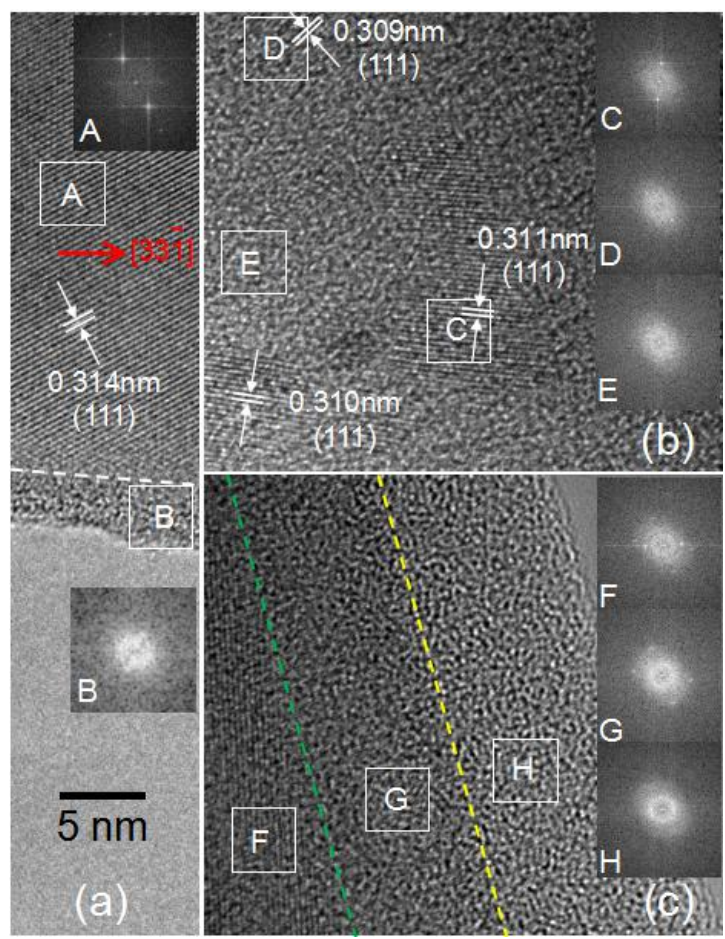

**Fig. 1** Representative HRTEM images showing (a) the typical Si nanowire structure (as indicated by Area 1) before irradiation as shown in A of Fig. S1(a), (b) the center (as indicated by Area 2) and (c) the surface (as indicated by Area 3) of the Si nanowire after 308 s irradiation as shown in I of Fig. S1(a). The insets are FFT images of their corresponding areas in HRTEM images.

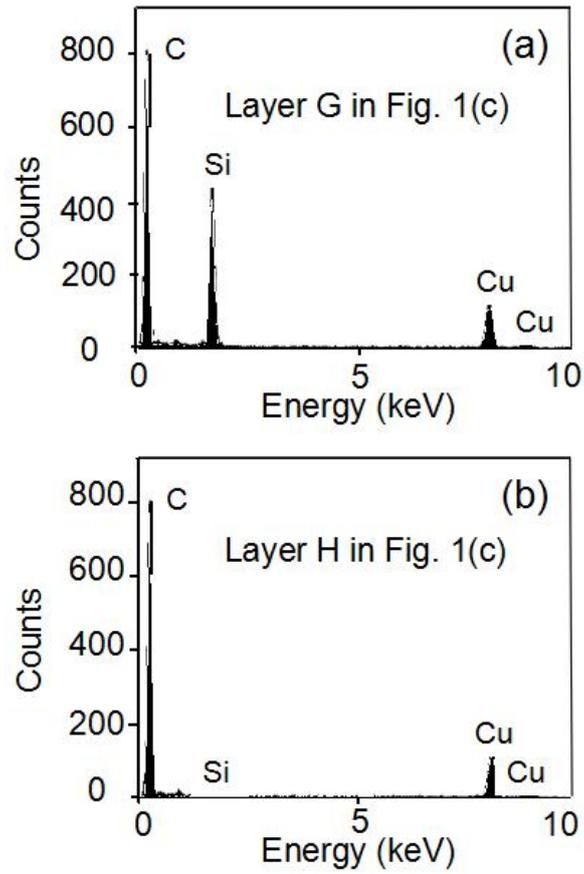

**Fig. 2** EDX spectra showing the compositions of layer G and H in Fig. 1(c).

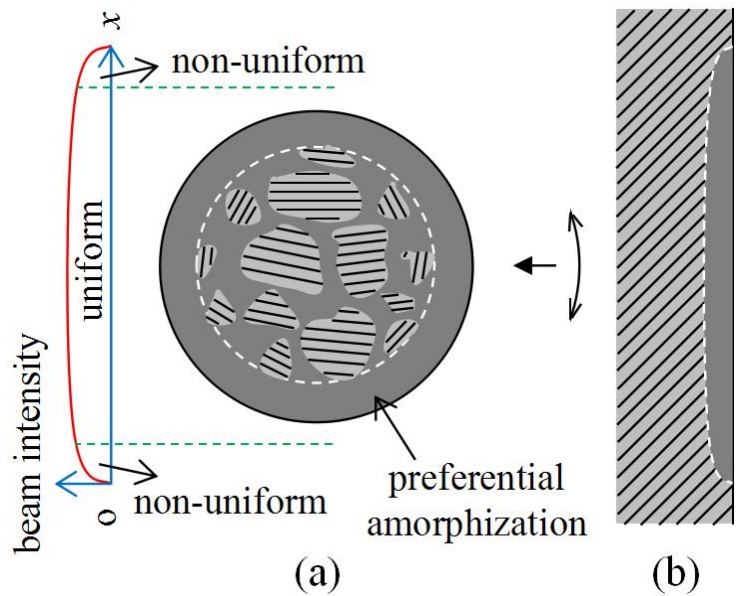

**Fig. 3** Schematic diagrams showing the structure difference and the resulting amorphization behavior between a nanowire (a) and a thin film (b) under focused electron beam irradiation.

16